\documentclass[twocolumn]{pasj00}

\SetRunningHead{S. Frey et al.}
{VSOP monitoring of the compact BL Lac object AO~0235+164}

\title{VSOP monitoring of the compact BL Lac object AO~0235+164}
\author{S\'andor \textsc{Frey}}
\affil{%
   F\"OMI Satellite Geodetic Observatory,
   P.O. Box 585, H-1592 Budapest, Hungary}
\email{frey@sgo.fomi.hu}

\author{Leonid I. \textsc{Gurvits}}
\affil{%
   Joint Institute for VLBI in Europe,
   Postbus 2, 7990 AA Dwingeloo, The Netherlands}

\author{Denise C. \textsc{Gabuzda}}
\affil{%
   Physics Department, University College Cork,
   Cork, Ireland}

\author{Chris J. \textsc{Salter},
        Daniel R. \textsc{Altschuler} and
        Phil \textsc{Perillat}}
\affil{%
   National Astronomy and Ionosphere Center,
   Arecibo Observatory, HC3 Box 53995, Arecibo, Puerto Rico 00612}

\author{Margo F. \textsc{Aller} and
        Hugh D. \textsc{Aller}}
\affil{%
   Astronomy Department, University of Michigan,
   Ann Arbor, MI 48109, USA}

\author{Hisashi \textsc{Hirabayashi}}
\affil{%
   Institite of Space and Astronautical Science, Japan Aerospace Exploration Agency\\
   3-1-1 Yoshinodai, Sagamihara, Kanagawa 229-8510, Japan}

\and

\author{Michael M. \textsc{Davis}}
\affil{%
   SETI Institute,
   515 N. Whisman Road, Mountain View, CA 94043, USA}

\Received{2005 June 30}
\Accepted{2005 September 12}
\Published{$\langle$publication date$\rangle$}
\begin{document}
\maketitle

\begin{abstract}
In 1999, the highly compact and variable BL Lac object AO~0235+164 was identified as the highest brightness temperature active galactic nucleus observed with the VLBI Space Observatory Programme (VSOP), with $T_{\rm B}>5.8\times10^{13}$~K \citep{fre00}.
The sub-milliarcsecond radio structure of this source has been studied with dual-frequency (1.6 and 5~GHz), polarization-sensitive VSOP observations during 2001 and 2002. Here we present the results of this monitoring campaign. At the time of these observations, the source was weakly polarized and characterized by a radio core that is clearly resolved on space--ground baselines.
\end{abstract}

\section{Introduction}
AO~0235+164 is an extensively studied BL Lac object at redshift $z=0.940$ \citep{coh87}.
The object is seen through a foreground group of galaxies at $z=0.524$ responsible for spectral lines in both absorption and emission. Absorption lines from $z=0.851$ are also present in the spectrum. The source is highly variable over the whole electromagnetic spectrum from radio to $\gamma$-rays. Variability time scales from years to days are observed, as well as intra-day variability. See e.g. \citet{rai01} and references therein for a detailed account of the optical and radio variability studies of AO~0235+164 over the past three decades.

The high-resolution radio structure of the source at milli-arcsecond (mas) angular scales revealed by ground-based Very Long Baseline Interferometry (VLBI) is characterized by a dominant compact core. Occasionally, faint extensions are seen at various position angles mainly between the north and east. A collection of references to earlier VLBI observations is given by \citet{fre00}. Based on a series of 43-GHz Very Long Baseline Array (VLBA) observations spanning almost two years, \citet{jor01} recently claimed to identify two components well within 1~mas of the core that show apparent superluminal motion with speeds up to $\beta_{\rm app}=30 h^{-1} c$. This would imply a Doppler factor of at least 90 in the jet when it points directly to the observer. Such high Doppler factors ($\sim100$) were reported by \citet{kra99} based on radio variability measurements, and by \citet{fuj99} derived from VLBI and radio total flux density observations. The result is also consistent with our earlier first-epoch 5-GHz VLBI Space Observatory Programme (VSOP) imaging observation \citep{fre00}. We derived a lower limit to the brightness temperature, $T_{\rm B}>5.8\times10^{13}$~K, based on an unresolved core component. This is the highest value measured directly and implies a Doppler factor of $\sim100$.

However, the Doppler boosting also varies with time as can be inferred from the brightness temperatures derived from VLBI data \citep{fre00}. This, and the large changes in the apparent VLBI jet position angle are qualitatively well explained by small deviations in a jet that intrinsically points very close to the line of sight. Indeed, \citet{ost04} interpreted the long-term quasi-periodic variability of AO~0235+164 at multiple wavebands as orientation variations in a helical jet. In their model, non-periodic features are explained by flow instabilities. Short timescale (intra-hour) variability data taken in 2000 are best interpreted by invoking interstellar scintillation of a source of at least 0.015~mas in angular size, implying a Doppler factor smaller than 40 in the relativistically moving component \citep{pen04}.

As opposed to other BL Lac objects, high-resolution radio polarization studies using ground-based VLBI at 5~GHz \citep{gab92} and 8.4~GHz \citep{gab96} found no convincing structure in either total or polarized intensity apart from the compact core of AO~0235+164.
Our VSOP monitoring observations presented here were aimed at investigating how the sub-mas scale radio structure of AO~0235+164 varies, and to see whether linear polarization structure can be detected with the superior angular resolution offered by VSOP at the frequencies of 1.6 and 5~GHz. The VSOP satellite, HALCA, receives only left-circularly polarized radiation, and has a limited sensitivity compared to ground-based VLBI stations. However, dual-polarization experiments with the co-observing ground network (e.g. the VLBA) allow polarization imaging of sources that show sufficiently high correlated polarized flux density \citep{kem00}.

   \begin{table}
      \caption{Space VLBI observations of AO 0235+164 at 1.6~GHz (top) and 5~GHz (bottom). Polarization-sensitive experiments are marked with asterisks}
         \label{observ}
         \begin{tabular}{llll}
            \hline
            \noalign{\smallskip}
             & Epoch      &  Ground network  & Correlator \\
            \noalign{\smallskip}
            \hline
            \noalign{\smallskip}
            A & 1999 Jan 31 & AT HH NO SH (4)  & Penticton \\
            B* & 2001 Feb 4  & VLBA AR (11)    & Socorro  \\
            C* & 2001 Aug 3  & VLBA AR GO (12) & Socorro  \\
            \noalign{\smallskip}
            \hline
            \noalign{\smallskip}
            D & 1999 Feb 1  & AT SH UD (3)     & Penticton \\
            E* & 2001 Feb 2  & VLBA Y (11)     & Socorro  \\
            F* & 2001 Aug 2  & VLBA AR (11)    & Socorro  \\
            G & 2001 Aug 14 & AR AT UD (3)     & Mitaka   \\
            H* & 2002 Jan 26 & VLBA (9)        & Socorro   \\
            \noalign{\smallskip}
            \hline
         \end{tabular}
	 Antenna names are AR: Arecibo (Puerto Rico); AT: Australia Telescope Compact Array;  GO: Goldstone (USA); HH: Hartebeesthoek (South Africa); NO: Noto (Italy); SH: Sheshan (China); UD: Usuda (Japan); VLBA: Very Long Baseline Array (USA);  Y: Very Large Array (VLA, USA)
   \end{table}

   \begin{figure}
   \begin{center}
   \FigureFile(8.5cm,6.0cm){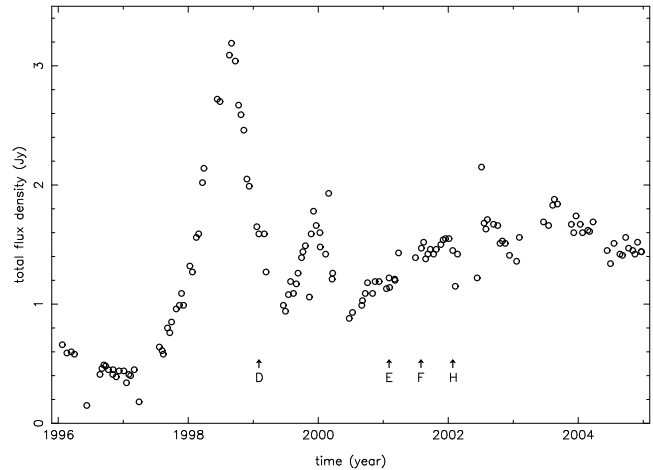}
   \caption{Total flux density history of AO 0235+164 at 4.8~GHz between 1996 and 2005. Two-week average UMRAO measurements are shown. Our SVLBI monitoring epochs are marked.}
         \label{fluxhistory}
   \end{center}
   \end{figure}

   \begin{figure}
   \begin{center}
   \FigureFile(6.5cm,6.5cm){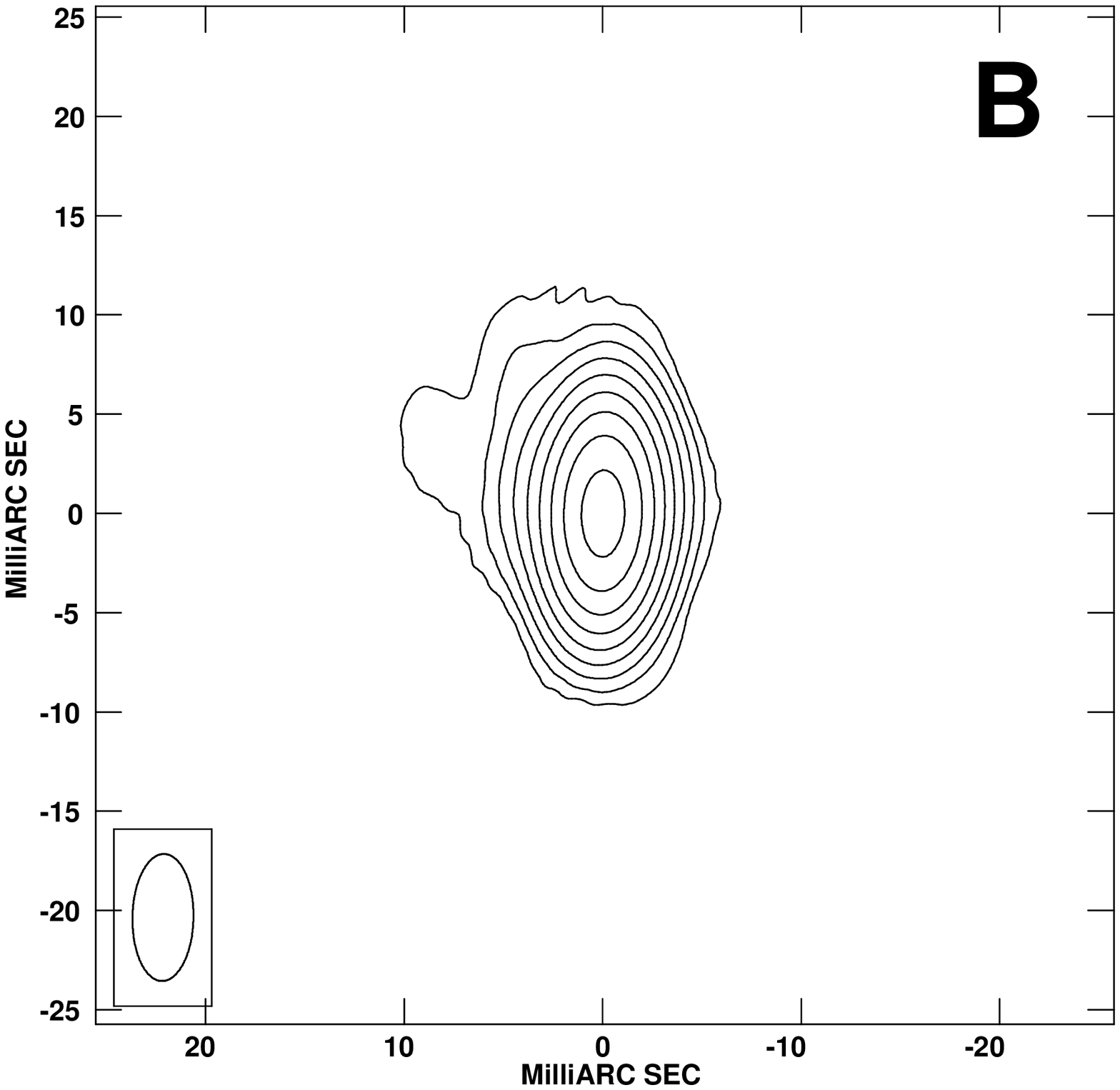}
   \FigureFile(6.5cm,6.5cm){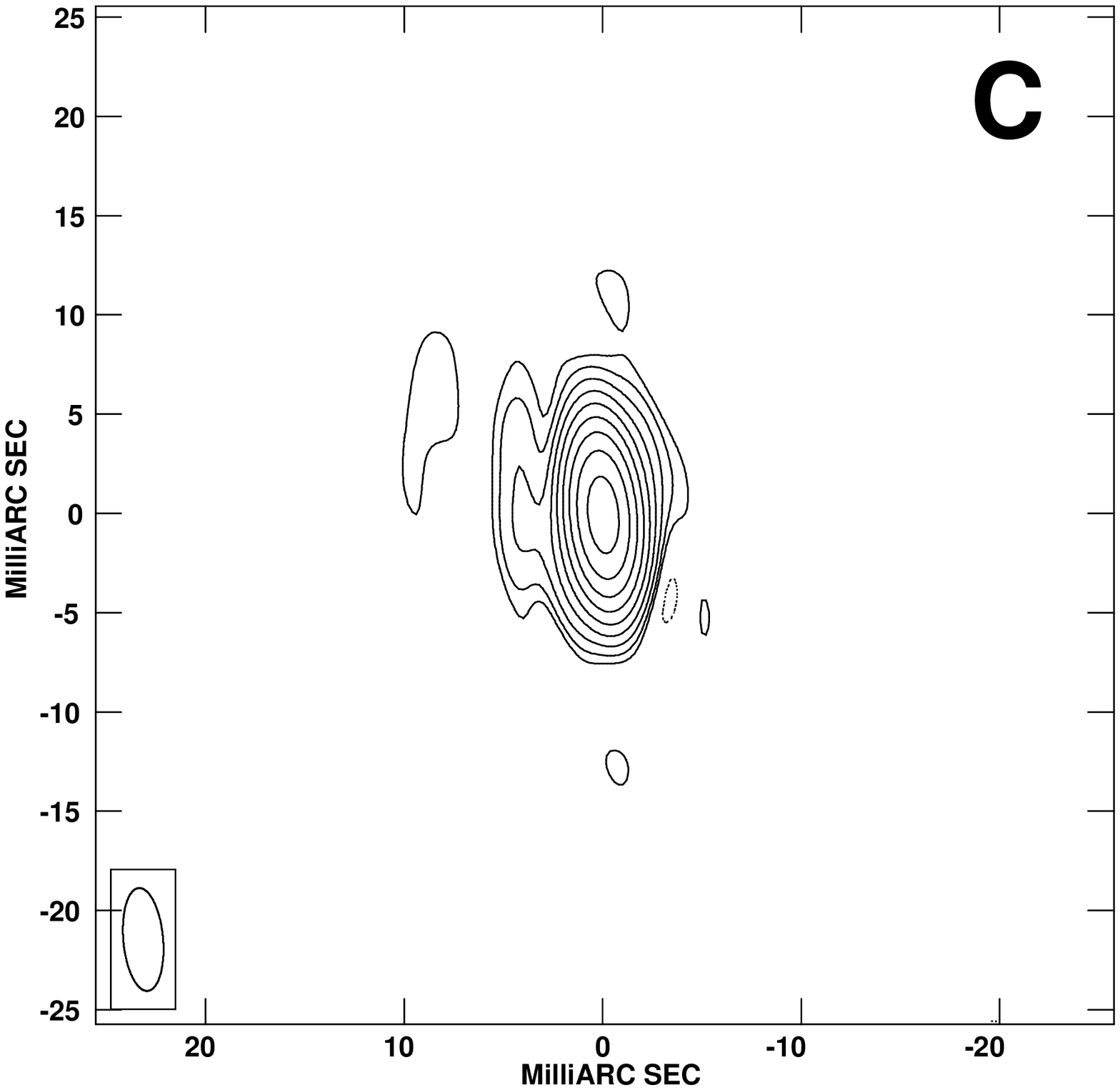}

      \caption{1.6-GHz SVLBI images of AO 0235+164 taken on 2001 Feb 4 and  2001 Aug 3 (epochs {\bf B} and {\bf C}). The first total intensity contours are drawn at $\pm2$~mJy/beam in each image. The positive contour levels
increase by a factor of 2. The peak brightnesses are 703 mJy/beam ({\bf B}) and 753 mJy/beam ({\bf C}). The restoring beams are 6.40~mas $\times$ 3.07~mas at the position angle (PA) $-1\arcdeg$ ({\bf B}) and 5.20~mas $\times$ 2.00~mas at PA=$5\arcdeg$ ({\bf C}), as displayed in the lower-left corners.
The off-source RMS image noise ($1\sigma$) is 0.4 mJy/beam ({\bf B}) and 0.5 mJy/beam ({\bf C}).
The right ascension (horizontal axis) and declination (vertical axis) scales have the brightness peak as origin.}
      \label{svlbi-image-1.6}
   \end{center}
   \end{figure}

\section{Observations and data reduction}

The details of the 3-year Space VLBI (SVLBI) monitoring program are summarized in Table~\ref{observ}. AO~0235+164 was observed at 1.6 and 5~GHz with VSOP at a total of 8 epochs, including the two experiments in 1999 (marked with A and D) already published by \citet{fre00}. Here we focus on the results of the other observations. For each experiment, the data from the space radio telescope on board the HALCA satellite \citep{hir00a} were recorded at a subset of the tracking stations at Goldstone, Green Bank (USA), Robledo (Spain), Tidbinbilla (Australia) and Usuda (Japan). The bandwidth was 32~MHz. As shown in Table~\ref{observ}, on different occasions, different VSOP data processors in Mitaka (Japan), Penticton (Canada) and Socorro (USA) were used. The total number of ground stations for each
experiment is given in parentheses in Table~\ref{observ}. Polarization-sensitive observations involving 10 (in one case 9) antennas of the VLBA are marked with asterisks after the experiment identifier in the first column.

Note that AO~0235+164 is included in the VSOP 5-GHz Active Galactic Nucleus (AGN) Survey sample \citep{hir00b}. However, due to its variability, at the time of the 5-GHz pre-launch VLBA observations \citep{fom00} it failed to reach the required correlated flux density for inclusion in the final source list to be observed with HALCA in the Survey.

At each observing epoch, multi-frequency measurements of the total radio flux density were obtained quasi-contemporaneously at the University of Michigan Radio Astronomy Observatory (UMRAO) and the Arecibo Observatory. AO~0235+164 is included in the long-term UMRAO monitoring program of total and polarized flux densities at 4.8, 8 and 14.5~GHz \citep{all99,all05}. An up-to-date 4.8-GHz total flux density curve covering our monitoring period is given in Fig.~\ref{fluxhistory}. See \citet{all05} for multiple frequencies and a longer time range. The single-dish flux density measurements were invaluable for verifying the calibration of our interferometer data.

The US National Radio Astronomy Observatory (NRAO) Astronomical Image Processing System (AIPS) was used for the data calibration and imaging. The visibility amplitudes were calibrated using system temperatures measured at the antennas wherever available. Nominal values were applied for HALCA and occasionally for some ground antennas (AR, AT and UD). Due to the sufficiently large ground network in most experiments, these could be verified and adjusted in the calibration procedure. The initial phase calibration was done ``manually'' using short scans of data.

\begin{figure}
   \begin{center}
   \FigureFile(6.5cm,6.5cm){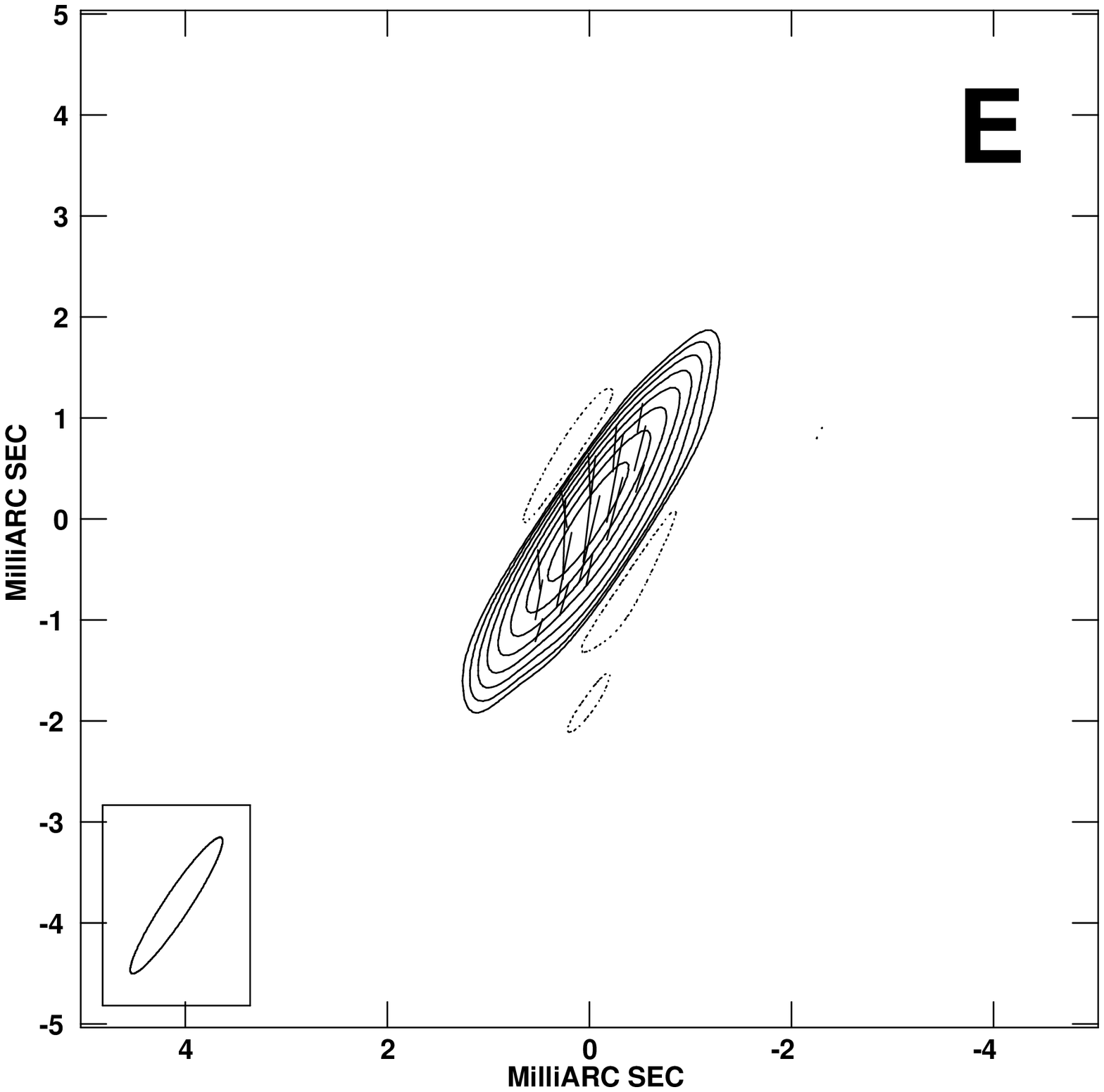}
   \FigureFile(6.5cm,6.5cm){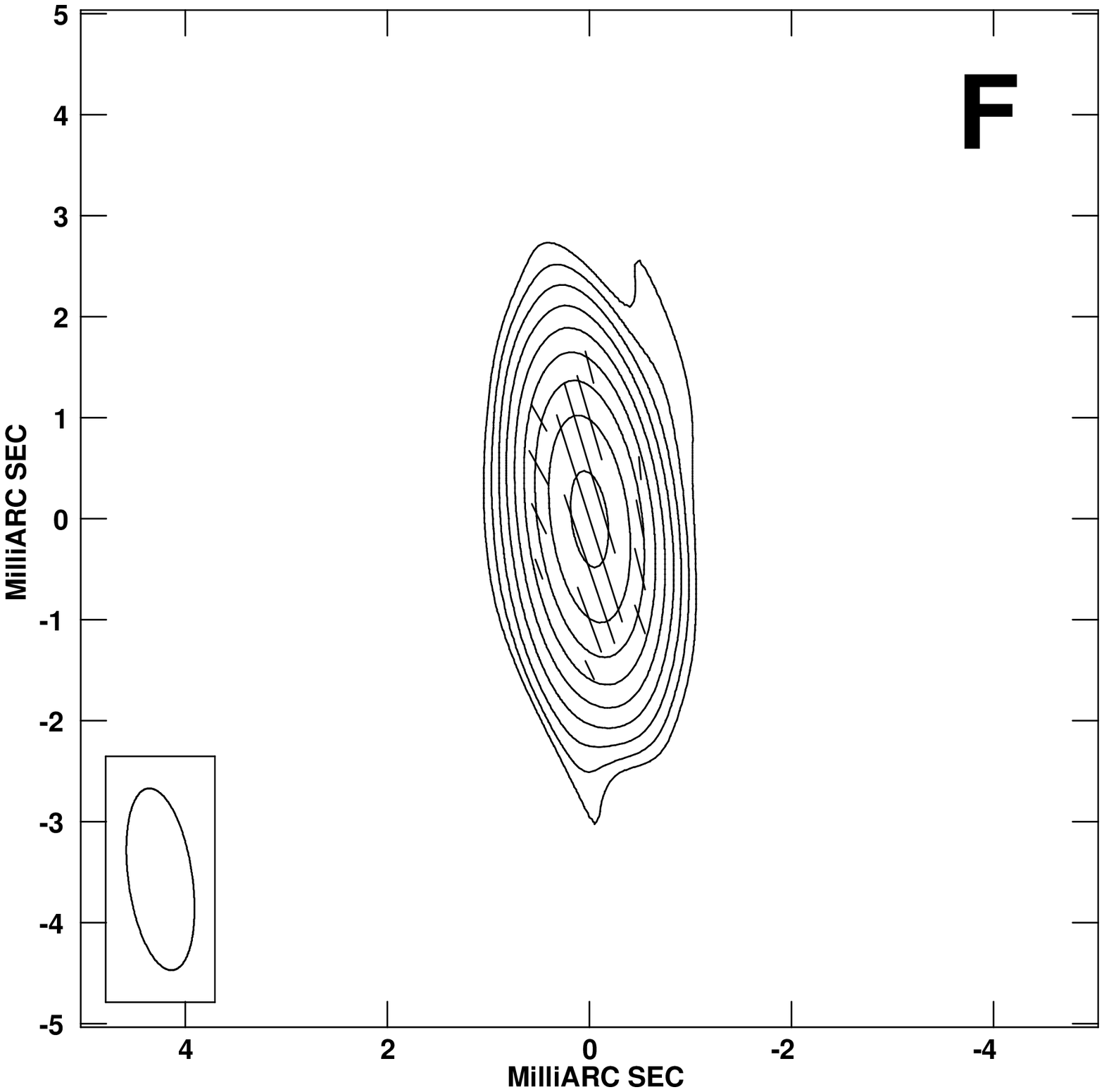}
   \FigureFile(6.5cm,6.5cm){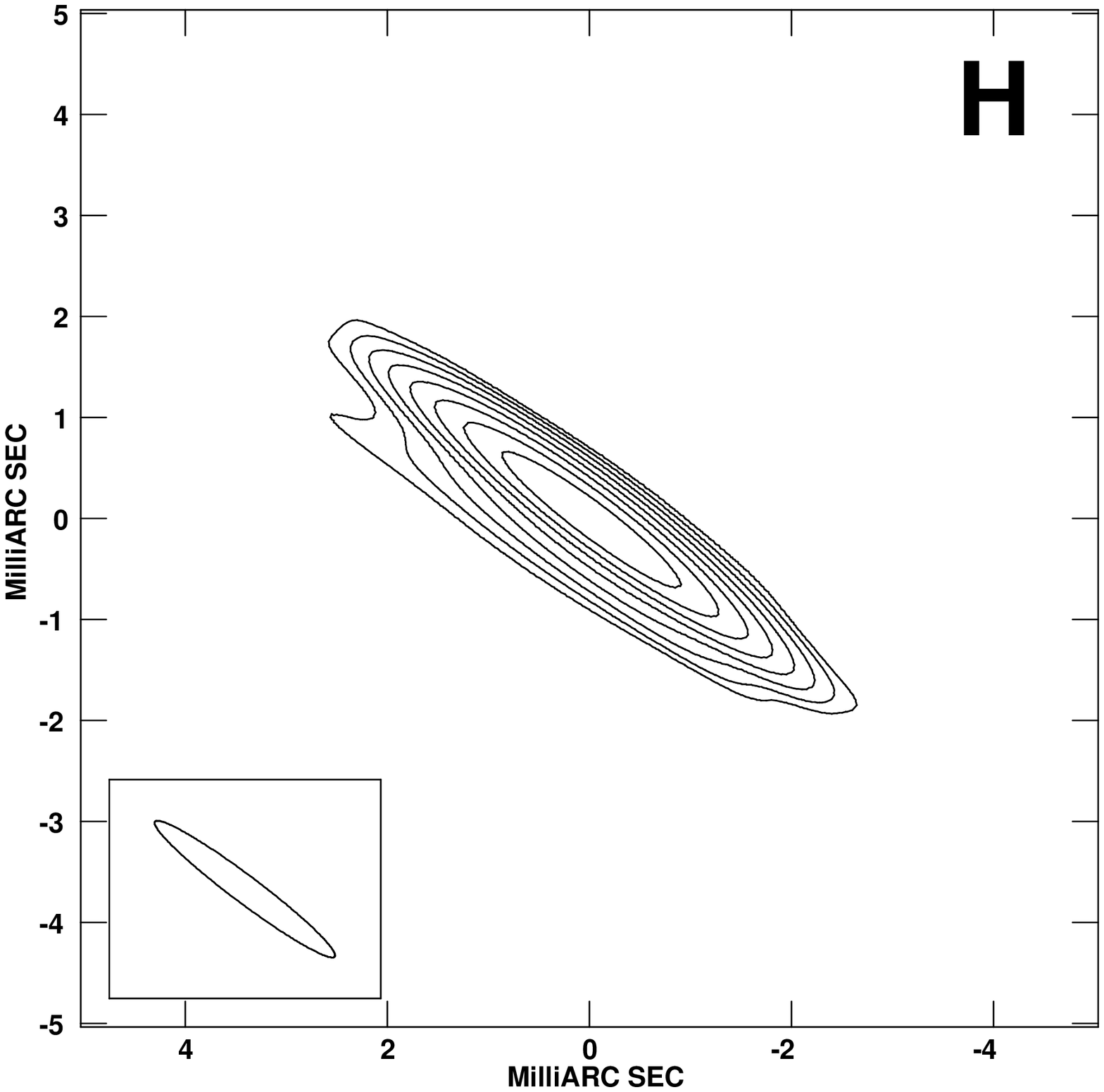}

      \caption{5-GHz SVLBI images of AO 0235+164 taken on 2001 Feb 2, 2001 Aug 2 and 2002 Jan 26 (epochs {\bf E}, {\bf F} and {\bf H}). The first total intensity contours are drawn at $\pm4$~mJy/beam in each image. The positive contour levels
increase by a factor of 2. The peak brightnesses are 859 mJy/beam ({\bf E}), 1243 mJy/beam ({\bf F}) and 1008 mJy/beam ({\bf H}). The restoring beams are 1.61~mas $\times$ 0.25~mas at PA=$-34\arcdeg$ ({\bf E}), 1.82~mas $\times$ 0.63~mas at PA=$8\arcdeg$ ({\bf F}) and 2.23~mas $\times$ 0.27~mas at PA=$53\arcdeg$ ({\bf H}), as displayed in the lower-left corners. The off-source RMS image noise ($1\sigma$) is 0.9 mJy/beam ({\bf E}, {\bf F}) and 1.1 mJy/beam ({\bf H}).
Electric vectors are superimposed ({\bf E} and {\bf F}), 1~mas corresponds to 10~mJy/beam polarized intensity.}
         \label{svlbi-image-5}
   \end{center}
   \end{figure}

In the case of the polarization-sensitive experiments (B, C, E, F and H), the right--left delay difference was removed with the AIPS procedure {\sc VLBACPOL}. Fringe-fitting was performed with the AIPS task {\sc FRING} over 5-minute solution intervals. The total intensity images are the results of a hybrid mapping procedure involving several iterations with the tasks {\sc CALIB} and {\sc IMAGR}. The fourth root of the visibility data weights were used in a uniform weighting scheme in order to increase the relative importance of the baselines to the orbiting radio telescope and hence to improve the angular resolution. The total intensity images are displayed in Fig.~\ref{svlbi-image-1.6} (1.6 GHz) and Fig.~\ref{svlbi-image-5} (5 GHz). Identical coordinate scales and contour levels are used for each epoch in both figures. No image is presented for the 5-GHz observation on 2001 Aug 14 (experiment G), because only single baselines were available for most of the time. However, the source was clearly detected on each baseline and the data are consistent with a slightly resolved compact component.

For the polarization experiments, the task {\sc LPCAL} was used to solve for instrumental polarization terms for the antennas. Absolute calibration of the polarization position angle was possible for experiments E and F. The calibrator source 0048-097 was observed with the ground antennas. Integrated polarization measurements for the calibrator were obtained from the NRAO Very Large Array (VLA) involved in the SVLBI experiment as a phased array (epoch E), and from the UMRAO monitoring (epoch F). In the latter case, due to a 2-month period between our epoch and the data base entry, the calibration is more uncertain. The corresponding electric vectors are shown in Fig.~\ref{svlbi-image-5}. The electric vector position angles are $-11\arcdeg$ and $34\arcdeg$ for experiments E and F, respectively.

While most of the ground antennas (the VLBA, the phased VLA) recorded in both right ($R$) and left ($L$) circular polarization, HALCA and some of the ground-based antennas received only left circular polarization. Therefore only $LL$ and $RL$ correlations were obtained for certain baselines including those to the orbiting antenna. In AIPS, the procedure {\sc CXPOLN} and the task {\sc CXCLN} were used to perform complex polarization imaging.

   \begin{figure}
   \begin{center}
   \FigureFile(6.5cm,6.5cm){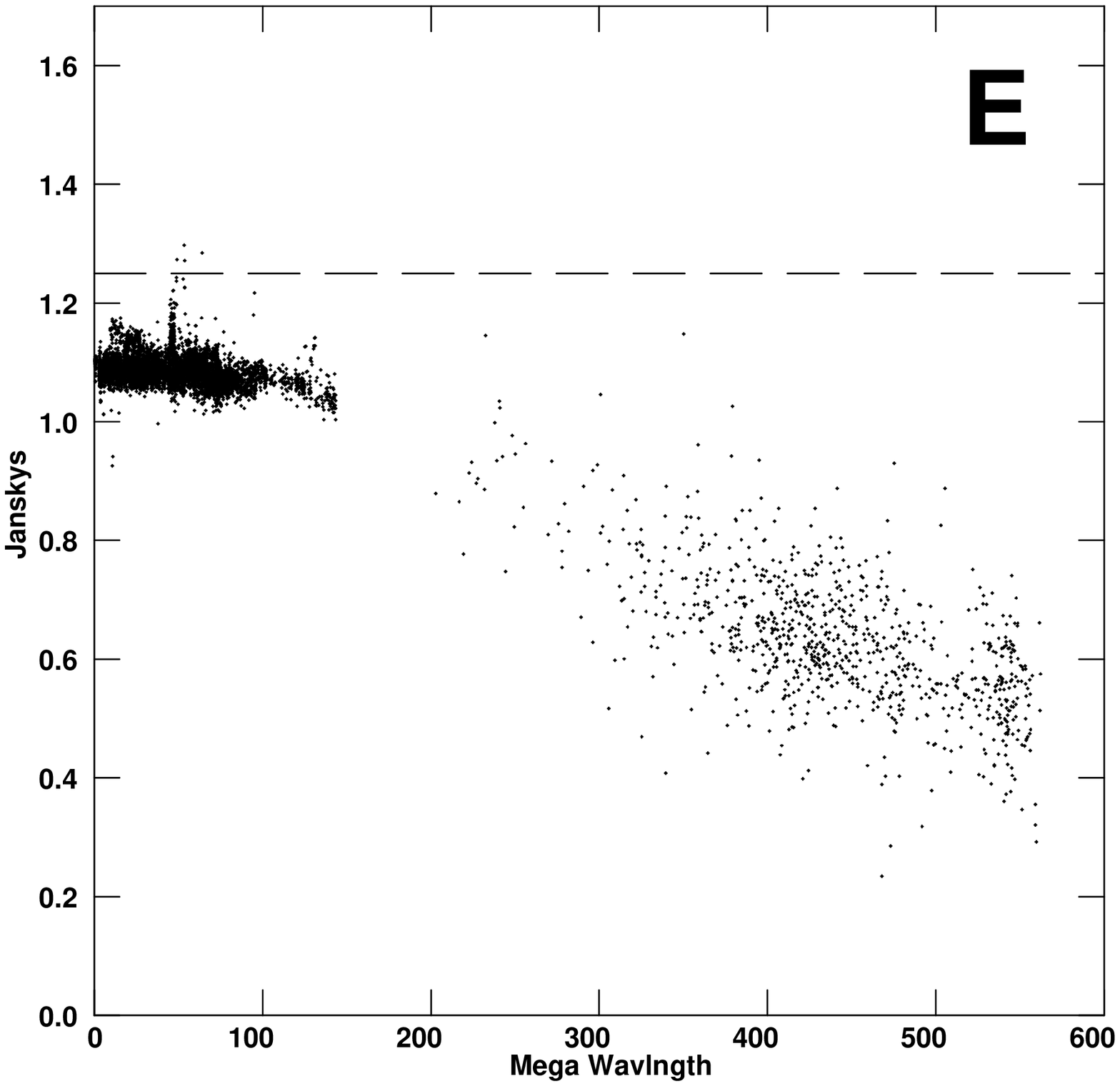}
   \FigureFile(6.5cm,6.5cm){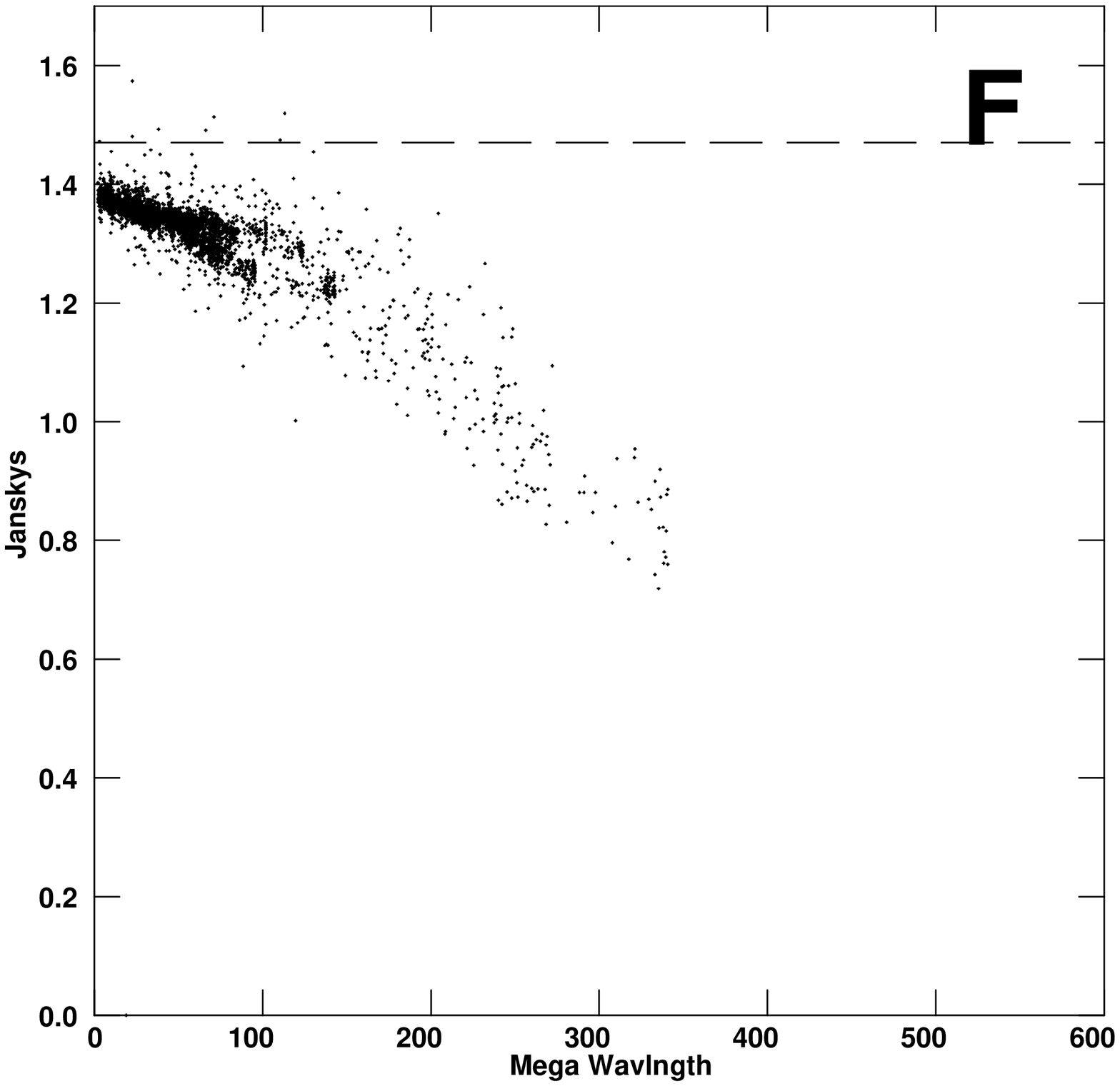}
   \FigureFile(6.5cm,6.5cm){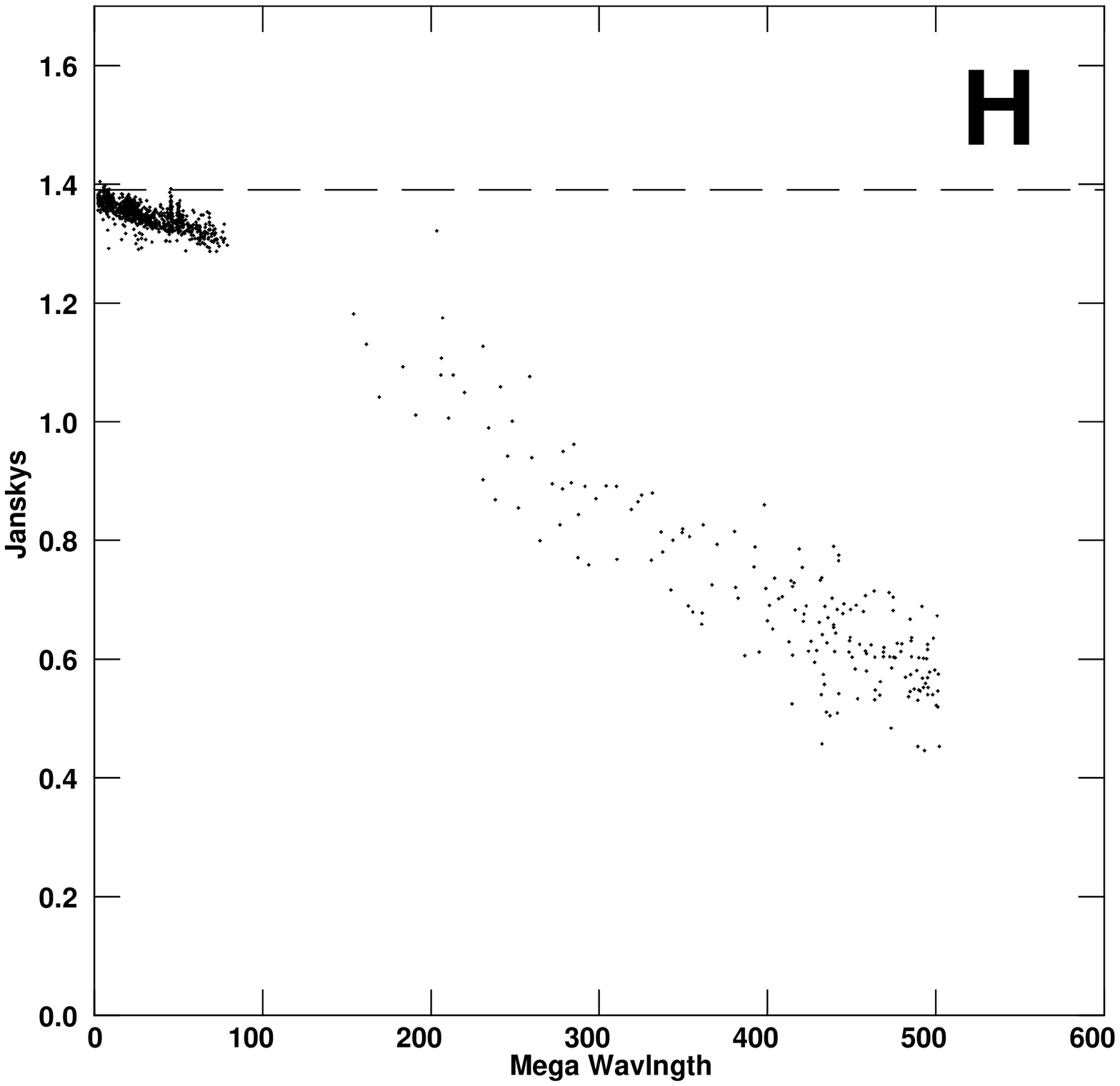}

      \caption{Self-calibrated 5-GHz correlated flux density vs. projected baseline length for the SVLBI observations of AO 0235+164 taken on 2001 Feb 2, 2001 Aug 2 and 2002 Jan 26 (epochs {\bf E}, {\bf F} and {\bf H}). Only every tenth point is plotted for clarity. Ground-only baselines typically extend up to $\sim150$ million wavelengths (M$\lambda$). Total (single-dish) flux density values obtained at the UMRAO at nearly the same time are indicated with horizontal dashed lines.
      }
         \label{svlbi-radpl-5}
   \end{center}
   \end{figure}

\section{Results and discussion}

At 1.6~GHz, the total intensity images of AO~0235+164 (Fig.~\ref{svlbi-image-1.6}) obtained at epochs B and C separated by 6 months in time show a resolved and somewhat extended structure. The low-brightness extension is on the eastern side of the compact core. The fractional polarization of the core was $\lesssim1$\%.

At 5~GHz, in contrast to our earlier results (epoch D, \cite{fre00}), the source is well resolved on the space--ground baselines at three different epochs. This is clearly indicated by the plots of the correlated flux density as a function of the projected baseline length (Fig.~\ref{svlbi-radpl-5}). Although there are hints of sub-mas scale structure in the images, these are difficult to interpret due to the highly elongated restoring beam, a consequence of the asymmetric ($u,v$)-coverage. In fact, the source brightness distribution is well fitted with a single circular Gaussian component of 0.20, 0.30 and 0.23~mas (FWHM) in the case of experiments E, F and H, respectively. These values imply brightness temperatures of $\sim10^{12}$~K, more than an order of magnitude less than that derived in 1999, shortly after a major total flux density outburst \citep{fre00}. The source has apparently become extended and, according to the UMRAO total flux density monitoring data, also faded since 1999. Our monitoring epochs fall into a long, relatively quiescent period in the flux density history of AO~0235+164 (Fig.~\ref{fluxhistory}).

The nearly contemporaneous 5-GHz total flux density measurements are indicated with horizontal dashed lines in the correlated flux density plots (Fig.~\ref{svlbi-radpl-5}) for comparison. It is interesting to note that the correlated flux densities on the shortest VLBI baselines are considerably lower than the total flux densities, by up to $\sim150$~mJy (epochs E, F). This  phenomenon is often encountered at different epochs and frequencies by other authors as well. In particular, \citet{ros04} lists AO~0235+164 as one of the ``compact but slightly resolved'' sources found in the VLBA 2-cm Survey sample, suggesting that a weak extended emission is associated with a jet structure seen more or less face-on. In the framework of this survey, \citet{kov05} derived $T_{\rm B}=9.08\times10^{11}$~K at 15~GHz at the epoch of 2001 Mar 15, consistently with our results discussed here.

   \begin{figure}
   \begin{center}
   \FigureFile(6.5cm,6.5cm){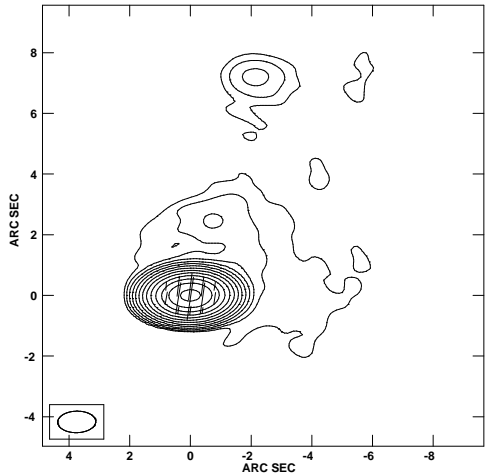}
        \caption{5-GHz image of AO 0235+164 made from the VLA data only (2001 Feb 2). The first total intensity contours are drawn at $\pm0.25$~mJy/beam. The positive contour levels increase by a factor of 2. The peak brightness is 1249 mJy/beam. The restoring beam is $1.25^{\prime\prime} \times 0.71^{\prime\prime}$ at PA=$-89\arcdeg$.
The off-source RMS image noise ($1\sigma$) is 0.05 mJy/beam.
Electric vectors are superimposed, $1^{\prime\prime}$ corresponds to 10~mJy/beam polarized intensity.}
         \label{vla-image}
   \end{center}
   \end{figure}

Since the phased VLA was included in the ground network in our 5-GHz experiment on 2001 Feb 2 (epoch E), we were also able to image AO~0235+164 using only VLA data. Our VLA image (Fig.~\ref{vla-image}) with about 3 orders of magnitude less resolution than that
of the SVLBI images shows weak extended emission surrounding the core, reaching 0.1\% of the peak brightness in a component at an angular distance of \timeform{7.5''} to the north-northwest. The structure seen is similar to earlier VLA results \citep{mur93}. Unlike the VLBI data, the VLA observations fully account for the total flux density. About 10~mJy flux density can be attributed to the extended structure in the VLA image. We believe that diffuse low surface brightness emission is present on intermediate angular scales, but is resolved out by VLBI.

As argued by e.g. \citet{fuj99}, the radio jet in AO~0235+164 may lie very close (within $1\arcdeg$) to the line of sight. It is possible that this angle changes with a period of $\sim5-6$ years \citep{rai01}, as in the helical jet model proposed by \citet{ost04}. However, the next major radio outburst predicted by the model for around early 2004 has yet to be detected \citep{rai05}.
Qualitatively, this explains a wide range of phenomena observed in the source, including (some) variability, rapid changes in the VLBI jet component position angles (when observed), and the diffuse radio emission surrounding the core at larger scales. In this picture, our SVLBI monitoring covers a period post-1999 that corresponds to a larger jet angle with respect to the line of sight, and therefore a less dominant Doppler boosting.
With the jet Doppler factor $\delta\sim100$ determined in 1999, the Lorentz factor characterising the bulk plasma motion is $\gamma \gtrsim 50$, and the corresponding jet viewing angle is $\theta \lesssim 0\fdg5$
\citep[Appendix A]{urr95}. At our later epochs, the Doppler boosting decreased by a factor of $\sim 20$. Under the simple assumption that the bulk Lorentz factor remained similar, the observed change in the brightness temperature can be explained by an increase in $\theta$ as small as $\lesssim 5\arcdeg$.

At our monitoring epochs, the fractional polarization of AO~0235+164 was quite low, $\sim0.5-2$\%. This result is fully consistent with the corresponding UMRAO monitorig data, where the polarized flux density was measured to be around the detection limit at these epochs. Historically, values as high as $\sim5$\% have been measured for AO~0235+164. Besides the low level of polarization, the structure found is simple, no feature apart from the compact core is being seen in our polarization images. This rather atypical property of AO~0235+164 compared to other BL Lac objects has already been noted by \citet{gab92} and \citet{gab96} using ground-based VLBI data. In this ``inactive'' phase the degree of ordering of the magnetic field in the source is low and no propagating shock is emerging in the plasma flow.

SVLBI observations of other BL Lac objects \citep[and references therein]{pus05} resolve the polarization structure of the inner jet components. This is typically aligned with the jet axis, suggesting that the jets are associated with helical magnetic fields that propogate outward with the jet flow.

\par
\vspace{1pc}\par
We gratefully acknowledge the VSOP Project, which is led by the Institute of Space and Astronautical Science (Japan) in cooperation with many organizations and radio telescopes around the world.
The National Radio Astronomy Observatory is a facility of the National Science Foundation (NSF) operated under cooperative agreement by Associated Universities, Inc.
The Arecibo Observatory is operated by Cornell University under cooperative agreement with the NSF. This research has made use of data from the University of Michigan Radio Astronomy Observatory which is supported by funds from the University of Michigan and the NSF.
SF acknowledges the Bolyai Research Scholarship received from the Hungarian Academy of Sciences. This work was partly supported by the Hungarian Scientific Research Fund (OTKA, grant no. T046097).

\end{document}